\documentclass[a4paper,12pt]{article}
\usepackage{multirow}
\usepackage{verbatim}
\usepackage{amsfonts}
\usepackage{mathrsfs}
\usepackage{amsmath}
\usepackage{amssymb}
\usepackage{float}
\usepackage{framed}
\usepackage[medium]{titlesec}
\usepackage{bm}
\usepackage{cite}
\usepackage[normalem]{ulem}
\usepackage{extarrows}
\usepackage{slashed}
\usepackage{isodateo}
\usepackage{graphicx}
\usepackage{xcolor}
\usepackage[bookmarksnumbered=true,bookmarksopen=true,hyperfootnotes=true]{hyperref}
\hypersetup{colorlinks,%
	linkcolor=[rgb]{0,0.3,0.6}, %
	citecolor=[rgb]{0,0.3,0.6}, %
	urlcolor=[rgb]{0,0.3,0.6}}
\usepackage[hmargin=.7in,vmargin=1.1in]{geometry}
\usepackage{indentfirst}
\usepackage{booktabs}
\usepackage{blindtext}
\usepackage{graphicx}
\usepackage{subcaption}
\usepackage[export]{adjustbox}
\usepackage{wrapfig}

\usepackage{enumerate}
\usepackage{amsfonts}
\usepackage{mathrsfs}
\usepackage{amsmath}
\usepackage{amssymb}
\usepackage{float}
\usepackage{bm}
\usepackage{slashed}
\usepackage{feynmp-auto}
\usepackage[normalem]{ulem}
\usepackage{extarrows}
\usepackage{slashed}
\usepackage{isodateo}
\usepackage{xcolor}
\usepackage{cancel}

\newcommand{\be}{\begin{equation}}
\newcommand{\ee}{\end{equation}}
\newcommand{\ba}{\begin{array}}
\newcommand{\ea}{\end{array}}
\newcommand{\bea}{\begin{eqnarray}}
\newcommand{\eea}{\end{eqnarray}}

\renewcommand{\d}{\mathrm{d}}

\newcommand{\besub}{\begin{subequations}}
\newcommand{\eesub}{\end{subequations}}

\newcommand{\nnn}{\nonumber \\}
\def\ma{\mathcal}

\def\fr{\frac}

\newcommand{\beq}{\begin{equation} \begin{aligned}}
		\newcommand{\eeq}{\end{aligned} \end{equation}}

\definecolor{darkerblue}{rgb}{0.2,0.2,0.5}
\definecolor{seagreen}{rgb}{0.180392,0.545098,0.341176}
\definecolor{smagenta}{rgb}{0.5,0.145098,0.341176}
\definecolor{deepblue}{rgb}{0,0,1}
\newcommand{\email}[1]{\footnote{Email: \href{mailto:#1}{\nolinkurl{#1}}}}

\usepackage{titlesec}
\usepackage{mdframed} 
\begin{document}

\title{\Large{\textbf{Large non-Gaussianities corresponding to first-order phase transitions during inflation}}}
\author{Haipeng An$^{1,2,\,}$\email{anhp@mail.tsinghua.edu.cn},~~ Qi Chen$^{1,\,}$\email{chenq20@mails.tsinghua.edu.cn},~~ Yuhang Li$^{1,\,}$\email{liyh19@mails.tsinghua.edu.cn}~~ and ~~Yuan Yin$^{1,\,}$\email{yiny18@mails.tsinghua.edu.cn}\\[5mm]
\normalsize{${}^{1}\,$\emph{Department of Physics, Tsinghua University, Beijing 100084, China}}\\ 
\normalsize{${}^{2}\,$\emph{Center for High Energy Physics, Tsinghua University, Beijing 100084, China}}\\ 
}


	
\date{}
\vspace{20mm}	
\maketitle
\begin{abstract}

In this study, we explore the back reaction of phase transitions in the spectator sector on the inflaton field during slow-roll inflation. Due to the significant excursion of the inflaton field, these phase transitions are likely to occur and can induce substantial non-Gaussian correlations in the curvature perturbation. Our results suggest that these correlations could be detectable by future observations of the cosmic microwave background radiation and large-scale structure surveys. Furthermore, we demonstrate that in certain parameter spaces, a scaling non-Gaussian signal can be produced, offering deeper insights into both the inflaton and spectator sectors. Additionally, phase transitions during inflation can generate gravitational wave signals with distinctive signatures, potentially explaining observations made by pulsar timing array experiments. The associated non-Gaussian correlations provide collateral evidence for these phase transitions.

\end{abstract}

\newpage
\tableofcontents	
\newpage
\section{Introduction}

It is highly plausible that the universe experienced an inflationary phase prior to the thermal Big Bang expansion~\cite{Starobinsky:1979ty,Starobinsky1980,Kazanas:1980tx,Sato:1981qmu,Guth:1980zm,Linde:1981mu,Albrecht:1982wi}. The most credible inflation models posit that the universe's accelerated expansion was driven by the potential energy of a slowly-rolling scalar field, with the field's excursion approaching or even larger than the Planck scale during the inflationary period. However, since inflation must eventually conclude and the substantial potential energy of the inflaton field must be transferred to the standard model thermal plasma to initiate the standard Big Bang expansion, the inflaton field must couple to other fields. Consequently, the significant excursion of the inflaton field may alter the properties of the fields coupled to it, potentially triggering a phase transition in the spectator sector~\cite{An:2020fff,An:2022cce,An:2024zfi,Tong:2023krn,Kumar:2021ffi,Bodas:2022urf,Bodas:2022zca}. In this context, we refer to the field that is coupled to the inflaton field but does not drive the universe's inflation as the spectator field. We hypothesize that the evolution of the inflaton field triggers a phase transition in the spectator field sector during inflation.

If the phase transition is first-order, it can generate gravitational waves through bubble collisions. The back reaction to the inflaton field will also produce significant curvature perturbations, which in turn will generate secondary gravitational waves upon re-entering the horizon after inflation~\cite{An:2023jxf}. These secondary gravitational waves can potentially account for the signals observed by pulsar timing array (PTA) collaborations~\cite{Xu:2023wog,NANOGrav:2023gor,Antoniadis:2023rey,Reardon:2023gzh}. Notably, the spectrum of these secondary gravitational waves aligns well with the power spectrum of the excess observed by NANOGrav~\cite{An:2023jxf}.

However, stochastic gravitational wave signals across most frequency ranges, once detected, may have multiple origins, spanning both astronomical and cosmological sources. These sources include supermassive black hole binaries, white dwarf backgrounds, cosmological phase transitions, and cosmic strings, among others (see~\cite{Caldwell:2022qsj} for a recent review). Therefore, it is crucial to identify additional signals that can help distinguish between the different origins of these gravitational wave sources. Fortunately, if a phase transition in the spectator sector occurred during inflation, the interactions within the spectator sector must have been sufficiently strong to temporarily enter a non-perturbative regime. In the symmetry-breaking phase, the homogeneous component of the spectator field $\sigma$, denoted as $\sigma_0$, evolves alongside the inflaton field $\phi$. Consequently, its perturbation $\delta\sigma$ also contributes to the curvature perturbation (see \cite{Gong:2016qmq} for a review of multi-field inflation models). Therefore, the interactions within the spectator sector during the symmetry-breaking phase induce a non-Gaussian correlation in the curvature perturbation. In this work, we investigate the non-Gaussian correlation function of the curvature perturbation associated with the phase transition in the spectator sector, triggered by the evolution of the inflaton field.


As will be demonstrated in Sections~\ref{sec:3pt} and \ref{sec:NG}, the non-Gaussianity, $f_{\rm NL}$, arising from the self-interaction of the $\sigma$ field benefits from several enhancement factors:
\begin{itemize}
    \item The three-point self-interaction in the spectator sector can be significantly larger than the Hubble scale, $H$. Consequently, as shown in Sec.~\ref{sec:3pt}, the induced 3pt interaction of the curvature perturbation ${\cal R}$ can be written as
    \bea
    {\cal C} \kappa^3 m_S^4 \epsilon^{3/2} {\cal R}^3 \ ,
    \eea
    where ${\cal C}$ is an order one coefficient, $\kappa \equiv M_{\rm pl}/\phi_0$ is typically order one in slow-roll inflation models, $m_S$ is the typical energy scale of the spectator sector, and $\epsilon$ is the slow-roll parameter. Consequently, compared to the cubic interaction of ${\cal R}$ in single field inflation model, the interaction is effectively enhanced by a factor of $\kappa^3 \epsilon^{-1/2} m_S^4/\rho_{\rm tot}$, where $\rho_{\rm tot}$ is the total energy density of the Universe during inflation. Since, in our model, the energy density is roughly entirely contributed by the vacuum energy driving inflation, we have $\rho_\text{tot}\simeq\rho_{\text{inf}}$. $m_S^4$ can also be estimated by $L$ the total latent heat released during inflation. Thus, compared to the single field inflation model, in our scenario $f_{\rm NL}$ is enhanced by a factor of 
    \bea
    \kappa^3 \frac{L}{\rho_\text{inf}} \epsilon^{-1/2} \ .
    \eea
    
    \item The three-point interaction of the curvature perturbation takes the form of ${\cal R}^3$. Consequently, the contribution to the non-Gaussianity does not diminish when the perturbation modes exit the horizon. This results in an enhancement of the time integral by $N_{\rm max}$, where $N_{\rm max}$ represents the number of e-folds between the phase transition and the point at which the hardest mode exits the horizon. In slow-roll inflation scenario, $N_{\rm max} \sim \epsilon^{-1/2}$. 
    
    \item The effective coupling of the ${\cal R}^3$ interaction is proportional to the time derivative of $\sigma_0$, which becomes singular as the system approaches the critical time of the phase transition. Thus, as discussed in Sec.~\ref{sec:3pt}, we expect an additional enhancement from the $\tau$ integral when $\tau$ approaches the critical point of the phase transition.

\end{itemize}

In summary, combining all the above enhancement factors, the resulted $f_{\rm NL}$ is about ${\cal O}(1)$.
Observations of the cosmic microwave background radiation (CMBR) have already constrained $f_{\rm NL}$ to be smaller than ${\cal O}(5)$~\cite{Planck:2019kim}. Future large-scale structure surveys may further probe $f_{\rm NL}$ down to the ${\cal O}(10^{-3})$ level~\cite{Euclid:2024yrr,DESI:2024mwx,SPHEREx:2014bgr,LSSTDarkEnergyScience:2012kar,PFSTeam:2012fqu,Eifler:2020vvg,Schlegel:2022vrv,Wang:2019jig,Liu:2022iyy}. 

Furthermore, the most significant contribution to the time integral of the bi-spectrum $B({\bf k}_1,{\bf k}_2,{\bf k}_3)$ of the curvature perturbation ${\cal R}$ arises from $|\tau| < k_{\rm max}^{-1}$, where $k_{\rm max}$ is the largest of $k_1$, $k_2$, and $k_3$. Consequently, we will demonstrate that the non-Gaussianity of ${\cal R}$ corresponds to the phase transition  scales with $k_{\rm max}$ slightly. This can be interpreted as a characteristic feature related to phase transition-induced non-Gaussian signals. 

In this work, we focus on the model where the phase transition induced by the evolution of the inflaton field results in symmetry restoration. For simplicity, we assume that the symmetry in question is a ${\cal Z}_2$ symmetry. By symmetry restoration, we mean that prior to the phase transition, the $\sigma$ sector is in the broken phase of the ${\cal Z}_2$ symmetry. Following a first-order phase transition induced by the inflaton evolution, the ${\cal Z}_2$ symmetry is restored, and $\sigma_0$ vanishes. Before the phase transition, since $\sigma_0\neq 0$, a three-point interaction of $\delta\sigma$ arises, which subsequently induces a three-point interaction of ${\cal R}$. After the phase transition, when $\sigma_0 = 0$, the three-point interaction of ${\cal R}$ generated by the self-interaction of $\sigma$ vanishes, thereby eliminating its contribution to non-Gaussianity. 


In our work, we utilize the symmetry-restoration phase transition as a key illustration primarily due to its relative simplicity and the insights it provides into the non-Gaussianity and gravitational wave signals generated during such transitions. By focusing on this specific type of phase transition, we can isolate the essential features of the gravitational waves and non-Gaussianities without the added complications that arise from symmetry-breaking scenarios.
In symmetry-breaking phase transitions, one encounters various topological defects—including domain walls, cosmic strings, and monopoles—that can significantly complicate the analysis. 
By avoiding these complexities, we ensure a clearer and more focused examination of how the symmetry restoration process influences the physical phenomena we are interested in.
Moreover, when we consider more intricate symmetries beyond the simplest scenarios, we find that the gravitational wave signals and non-Gaussianities of the symmetry restoration phase transitions exhibit similarities to the well-studied ${\cal Z}_2$ case. This resemblance enables us to extend our results and draw broader conclusions applicable to more complex symmetry systems without getting bogged down by the intricacies associated with symmetry-breaking phenomena. Consequently, this approach allows us to provide a more coherent and comprehensive understanding of the relationship between symmetry restoration processes, non-Gaussianity, and gravitational wave emissions in the context of cosmological phase transitions.

As demonstrated in \cite{An:2020fff,An:2022cce,An:2023jxf}, the strengths of the primary and secondary GW signals are proportional to $(\beta/H)^{-5}$ and $(\beta/H)^{-6}$, respectively, where $H$ is the Hubble expansion rate during inflation and $\beta$ is the rate of change of the bounce action. In contrast, the size of the non-Gaussianity parameter $f_{\rm NL}$ is proportional to $(\beta/H)^3$. The value of $\beta/H$ is influenced by the specifics of the inflaton sector, the spectator sector, and their interactions. Consequently, its value can be arbitrary. This variability indicates that gravitational waves and non-Gaussianity are complementary tools for exploring phase transition models in cosmology, as they provide different yet interconnected insights into the underlying dynamics.
 
The rest of the paper is organized as follows. In Sec.~\ref{sec:model}, we review the inflaton-driven phase transition models. Sec.~\ref{sec:PT} provides a detailed discussion of the symmetry restoration phase transition, where we calculate the relationship between the bounce action and the latent energy density released during the phase transition using a minimal polynomial spectator model. In Sec.~\ref{sec:3pt}, we evaluate the three-point interaction of curvature perturbations induced by the self-interaction of the $\sigma$ field, demonstrating that it is enhanced compared to single-field contributions as shown in~\cite{Maldacena:2002vr}. Sec.~\ref{sec:NG} presents our calculation of $f_{\rm NL}$ in the minimal phase transition model with a polynomial potential, discussing both its size and scaling behavior. In Sec.~\ref{sec:NGvsGW}, we examine the complementarity between non-Gaussianity and gravitational wave signals. While the main text focuses on the $\sigma$ sector with a polynomial potential, we note that if the field value is large during the phase transition, the potential may take the form of a Coleman-Weinberg (CW) potential, rendering a polynomial expansion inadequate. Therefore, in Sec.~\ref{sec:CW}, we discuss $f_{\rm NL}$ and gravitational wave signals assuming a CW type potential for the $\sigma$ sector. We conclude with a summary in Sec.~\ref{sec:summary}.

\section{Symmetry restoration phase transition induced by inflaton evolution}
\label{sec:model}

In this work, we consider the interaction between the inflaton field $\phi$ and a spectator field $\sigma$ through a direct coupling term of the form $c \phi^2 \sigma^2$. In standard inflationary models, it is typical to introduce derivative interactions between these fields, as such interactions do not induce quantum corrections to the inflaton mass. However, given that quantum gravity effects are anticipated to violate all continuous global symmetries~\cite{Kawasaki:2000ws,Adams:2006sv,Baumann:2014nda,Brennan:2017rbf,Harlow:2022ich}, it becomes plausible to include direct coupling terms between the inflaton and spectator fields. To prevent the $\eta$ problem \cite{Kawasaki:2000ws,Baumann:2014nda}, it is necessary that the coupling constant $c$ is constrained to be of the same order as the slow-roll parameters during inflation. Our hypothesis posits that the evolution of the inflaton field throughout the inflationary epoch instigates a symmetry-restoration phase transition by modifying the mass squared of the spectator field $\sigma$. In this context, $\sigma$ can also be interpreted as an order parameter for the phase transition.
We specifically focus on a phase transition consistent with a ${\cal Z}_2$ symmetry.
Consequently, the corresponding potential for the $\sigma$ field takes the form:
\bea\label{eq:V}
V(\sigma) = -\frac{1}{2}c \phi^2 \sigma^2 + V_0(\sigma) \ ,
\eea
where $V_0(\sigma)$ is independent of $\phi$. Thus, the mass term of the $\sigma$ field in the potential can be expressed as:
\bea\label{eq:Vc}
\frac{1}{2} (m_0^2 - c\phi_0^2) \sigma^2 \ ,
\eea
where $\phi_0$ is the homogeneous part of the inflaton field. We assume both $m_0^2$ and $c$ are positive parameters. As a result, the evolution of the homogeneous part of the inflaton field $\phi_0$ will lead to a scenario where the mass squared of the $\sigma$ field shifts from positive to negative values, thereby catalyzing a phase transition. 

For a phase transition to be complete, the occupation fraction of the true vacuum must approach ${\cal O}(1)$.  
The bubble nucleation rate per unit physical volume can be expressed as:
\bea
\frac{\Gamma}{V_{\rm phy}} = C m_\sigma^4 e^{-S_b} \ ,
\eea
where $S_b$ is the bounce action, $m_\sigma$ is the typical energy scale of the spectator sector, and $C$ is an order one parameter. Thus, at time $t$ the bubble nucleation rate per unit comoving volume can be written as
\bea
\frac{\Gamma}{V} = e^{3 Ht} C m_\sigma^4 e^{- S_b} \ .
\eea
For bubbles produced by vacuum phase transition, their surfaces expand at the speed of light, and thus for a bubble nucleated at $t'$, its comoving radius at $t$ can be written as 
\bea
R(t,t') = \frac{1}{H} \left(  e^{- H t'} - e^{-H t} \right) \ .
\eea
Then the fraction of the universe that remains in the false vacuum at time $t$ can be estimated as~\cite{Guth:1981uk}
\bea
{\cal P}(t) = \exp\left[ - \int_{-\infty}^t dt' \frac{4\pi}{3} R^3(t,t') \frac{\Gamma(t')}{V} \right] \ .
\eea
Therefore, for the phase transition to be complete, we have 
\bea\label{eq:requirement}
\int_{-\infty}^t dt' \frac{4\pi}{3} R^3(t,t') \frac{\Gamma(t')}{V} = {\cal O}(1) \ .
\eea
During the phase transition, we can expand $S_b$ around time $t$ to get
\bea
S_b(t') \approx S_b(t) + \frac{dS_b(t)}{dt}(t'-t) \equiv S_b(t) - \beta(t' - t) \ .
\eea
For the phase transition to be complete within a Hubble time, we usually require $\beta\gg H$. During the phase transition, $\beta$ can be estimated as a constant. Then the requirement (\ref{eq:requirement}) can be written as
\bea
8\pi C e^{- S_b(t)} \frac{m_\sigma^4}{\beta^4} = {\cal O}(1) \ .
\eea
Thus, at the time, $t_c$ when the phase transition is about to be complete, we have
\bea\label{eq:Sb1}
S_b(t_c) \approx \log\left(\frac{m_\sigma^4}{\beta^4} \right) \ .
\eea
We require only a small portion of the total energy density to be released during the phase transition so that the phase transition does not terminate the inflation. Furthermore, as discussed in \cite{An:2020fff,An:2022cce,An:2022toi}, for first-order phase transition occurring during inflation, it is common to have $\beta/H = {\cal O}(10)$. Consequently, from the previous equation, we get
\bea
S_b(t_c) \lesssim  \log\left( \frac{\rho_{\rm inf}}{H^4} \right) \ \approx 2 \log\left( \frac{M_{\rm pl}}{H} \right) \ .
\eea
Thus, for high-scale inflation models where $H \sim 10^{14}~{\rm GeV}$, we get $S_b \sim 20$, and for low-scale inflation models, $S_b$ can be about 200 (assuming reheating temperature to be about 1 MeV).

\section{General properties of the bounce action}
\label{sec:PT}

During a first-order phase transition characterized by symmetry breaking or restoration, the potential curve must exhibit at least three local minima and two local maxima. Consequently, the potential can be generically approximated during the phase transition by the integral:
\bea\label{eq:Vsigma}
V(\sigma) = \int d\sigma A \sigma (\sigma^2 - v_1^2) (\sigma^2 - v_2^2) \ .
\eea
During the phase transition, the energy scale of the tunneling process is significantly greater than the Hubble expansion rate of the universe. Hence, when evaluating the bubble nucleation rate, we can safely neglect the expansion of the universe. The Euclidean action can then be expressed as: 
\bea
{\cal S}_E = \int d^4 x_E \left[\frac{1}{2}\partial_\mu \sigma \partial_\mu\sigma + V(\sigma) \right]  \ . 
\eea
The bounce action, which determines the nucleation rate, is derived from this expression. Notably, the action ${\cal S}_E$ is dimensionless. 
To calculate ${\cal S}_E$, we can redefine $x_E$ and $\sigma$ in a manner that absorbs the dimensions present in the potential $V(\sigma)$. We define
\bea\label{eq:Delta0}
v_0^2 = v_1^2 + v_2^2 \ , \;\; \xi = \frac{\sigma}{v_0} \ , \;\; z= x v_0 \ , \;\; \frac{v_2}{v_1} = 1 + \Delta \ .
\eea
By comparing Eq.~(\ref{eq:Vsigma}) to (\ref{eq:Vc}), it is evident that both $A$ and $v_0$ remain constant throughout the evolution of $\phi_0$. Therefore, when we parameterize $V(\sigma)$ using $A$, $v_0$, and $\Delta$, only $\Delta$ varies with $\phi_0$. 

\begin{figure}[t]
 \centering
   \includegraphics[width=0.7\textwidth]{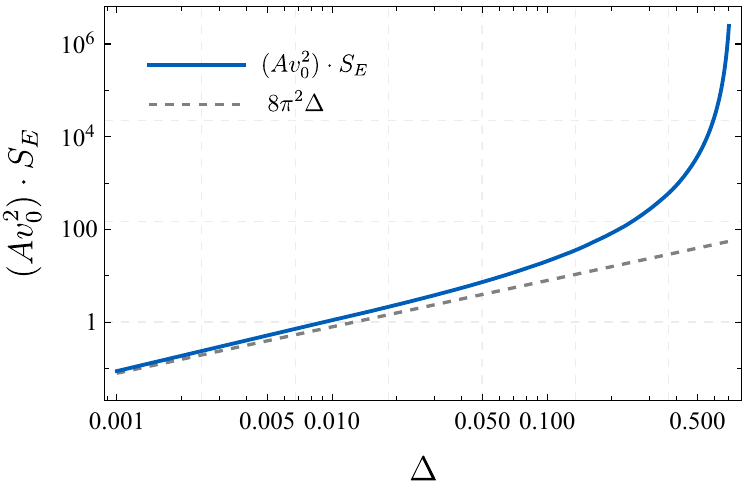}
\caption{$(Av_0^2) S_E$ as a function of $\Delta$.}
\label{fig:S4}
\end{figure}

With the redefined variables, the Euclidean action can be expressed as
\bea
{\cal S}_E = \int d^4 z \left[ \frac{1}{2} \frac{\partial\xi}{\partial z_\mu} \frac{\partial\xi}{\partial z_\mu} + Av_0^2 \tilde V(\xi) \right] \ ,
\eea
where 
\bea
\tilde V(\xi) = \frac{1}{6}\xi^6 - \frac{1}{4} \xi^4 + \frac{1}{2} \frac{(1+\Delta)^2}{[1+(1+\Delta)^2]^2} \xi^2 \ .
\eea
From the form of $\tilde V$ we can easily ascertain that $\xi = 0$ is a stable minimum when $\Delta < \sqrt{3}-1$, and meta-stable otherwise. 
Now, define $z = (A^{1/2} v_0) y$, we have
\bea\label{eq:SE1}
{\cal S}_E = (A v_0^2)^{-1} \hat {\cal S}_0(\Delta) \ , 
\eea
where
\bea
\hat{\cal S} = \int d^4 y \left[ \frac{1}{2} \frac{\partial\xi}{\partial y_\mu} \frac{\partial\xi}{\partial y_\mu} + \tilde V(\xi) \right] \ .
\eea
$\hat{\cal S}$ depends solely on $\Delta$, and its numerical value can be computed using CosmoTransitions~\cite{Wainwright:2011kj}. The results are illustrated in Fig.~\ref{fig:S4}. In the region where $\Delta\ll1$, $\hat{\cal S}$ exhibits a linear dependence on $\Delta$, with a coefficient of approximately $10\pi^2$. As $\Delta$ approaches $\sqrt{3}-1$, the thin wall approximation becomes applicable, leading to 
\bea
\hat{\cal S} \rightarrow 27\pi^2 \left(\frac{\sqrt{3}-1}{\sqrt{3}-1 - \Delta}\right)^3  \ .
\eea
It turns out that the following empirical formula 
\bea
\hat{\cal S} \approx 10\pi^2  \left[1.85 + 0.85\frac{(2\pi\Delta)^2 - 1}{(2\pi\Delta)^2 + 1 }\right] \left(\frac{\sqrt{3}-1}{\sqrt{3}-1-\Delta}\right)^3
\eea
agrees with the numerical result within 20\%.   

From Eq.~(\ref{eq:Vsigma}), we can calculate the latent energy density released during the phase transition, expressed as
\bea\label{eq:latent}
L = \frac{A v_0^6}{12} \frac{(1+\Delta)^4 (3 - (1+\Delta)^2)}{[1+(1+\Delta)^2]^3} \ .
\eea
In the limit where $\Delta \ll 1$, we find
\bea
L \approx \frac{1}{48} A v_0^6 \ .
\eea

\section{3pt interaction of the curvature perturbation in the symmetry breaking phase}
\label{sec:3pt}

\begin{figure}[t]
 \centering
   \includegraphics[width=0.45\textwidth]{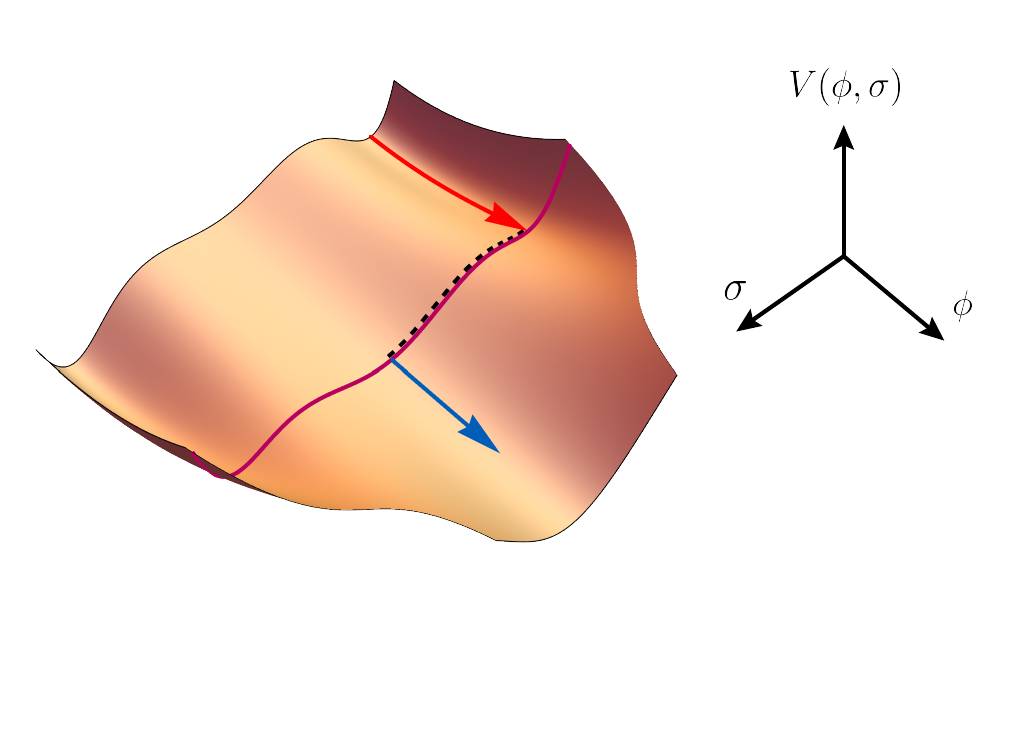}
   \includegraphics[width=0.45\textwidth]{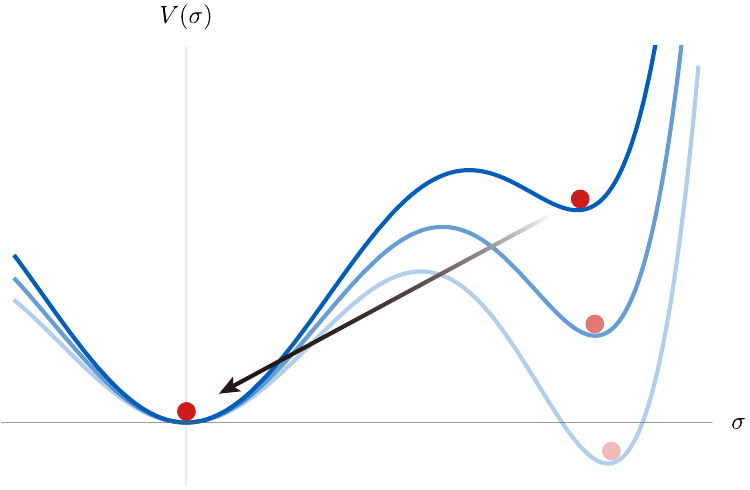}
\caption{An illustration of the potential of the spectator sector and the trajectory of the background evolution where the dashed curve indicates the tunneling of the meta-stable state.}
\label{fig:illustrate}
\end{figure}

As illustrated in Fig.~\ref{fig:illustrate}, during the evolution of $\phi_0$, $\sigma_0$, also undergoes changes. Consequently, both $\delta\phi$ and $\delta\sigma$ contribute to the curvature perturbation. While we will not provide a detailed derivation here, we will present the key results. For those interested in the complete derivation, we recommend consulting the relevant literature~\cite{Gong:2011uw,Achucarro:2012sm,Gong:2016qmq}.

We project the perturbations in the tangential and normal directions to the trajectory of the field evolution, yielding:
\bea
\delta \phi^{T} = &\frac{\dot{\phi}_0}{\sqrt{\dot{\phi}_0^2 + \dot{\sigma}_0^2}} \delta \phi + \frac{\dot{\sigma}_0}{\sqrt{\dot{\phi}_0^2 + \dot{\sigma}_0^2}} \delta \sigma  \ , \\
\delta \phi^N = &- \frac{\dot{\sigma}_0}{\sqrt{\dot{\phi}_0^2 + \dot{\sigma}_0^2}} \delta \phi + \frac{\dot{\phi}_0}{\sqrt{\dot{\phi}_0^2 + \dot{\sigma}_0^2}} \delta \sigma  \ .
\eea
This projection is universal, and when $\dot\sigma_0 = 0$, it simplifies to the identity matrix. Consequently, our expression reduces to the standard perturbation formalism. In the comoving gauge, the curvature and isocurvature fluctuations can be expressed as follows: 
\bea
\ma{R}= -\fr{H}{\sqrt{\dot{\phi}_0^2 + \dot{\sigma}_0^2}} \delta \phi^T = - H \left[\frac{\dot\phi_0 \delta\phi + \dot\sigma_0 \delta\sigma}{\dot{\phi}_0^2 + \dot{\sigma_0^2}}\right] \ , \;\; \ma{F} = \delta \phi^N  \ .
\eea
Since the Lagrangian is formulated in terms of the fields $\delta\phi$ and $\delta\sigma$ rather than curvature and isocurvature perturbations, it is more convenient for subsequent calculations to express the field fluctuations in terms of curvature and isocurvature perturbations:
\bea
\delta\phi &=& -\frac{\dot\phi_0}{H}{\cal R} -  \frac{\dot\sigma_0}{(\dot\phi_0^2 +\dot\sigma_0^2)^{1/2}}{\cal F}, \nnn
\delta\sigma &=& - \frac{\dot\sigma_0}{H}{\cal R} + \frac{\dot\phi_0}{(\dot\phi_0^2+\dot\sigma_0^2)^{1/2}} {\cal F} \ .
\eea
Consequently, the interaction terms in the scalar potential $V_\sigma$ leads to a three-point interaction of $\mathcal{R}$~\cite{Gong:2016qmq}:
\bea\label{eq:3pt}
- \left[\frac{1}{2}\frac{\partial^3 V}{\partial\phi^2\partial\sigma} \frac{\dot\phi_0^2\dot\sigma_0}{H^3} + \frac{1}{2} \frac{\partial^3 V}{\partial\phi\partial\sigma^2} \frac{\dot\phi_0\dot\sigma_0^2}{H^3} + \frac{1}{6}\frac{\partial^3 V}{\partial\sigma^3} \frac{\dot\sigma_0^3}{H^3}\right]{\cal R}^3 \ .
\eea
Assuming a power law interaction between $\phi$ and $\sigma$ as given in (\ref{eq:V}), we can estimation the derivatives of $V$ with respect to $\phi_0$ and $\sigma_0$:
\bea\label{eq:eta1}
\frac{\partial^3V}{\partial\phi^2\partial\sigma} = -\frac{\eta_1 \kappa^2 m_S^3}{M_{\rm pl}^2} \ , 
\eea
where $m_S$ is the typical energy scale of the spectator sector around the phase transition, $\kappa \equiv M_{\rm pl}/\phi_0$, and $\eta_1$ is a dimensionless parameter that is naturally of order one. Similarly, the second term in the square bracket in (\ref{eq:3pt}) can be parameterized as 
\bea\label{eq:eta2}
\frac{\partial^3 V}{\partial\phi \partial\sigma^2} = -\frac{\eta_2 \kappa m_S^2}{M_{\rm pl}} \ .
\eea
For the third term, we have
\bea\label{eq:eta3}
\frac{\partial^3V}{\partial\sigma^3} = \eta_3 m_S \ .
\eea

Another clear contribution to the ${\cal R}^3$ interaction comes from $\partial^3 V/\partial\phi^3$. However, the dominant part of this contribution arises from single-field-like interactions and is negligible due to Maldacena's theorem~\cite{Maldacena:2002vr}. The only opportunity in this model to obtain large non-Gaussianity is by considering interactions with the spectator sector. 

To calculate $\dot\sigma_0$, we note that $\sigma_0$ satisfies the equation $\partial V/\partial\sigma=0$. Therefore, we have
\bea\label{eq:sigmadotpre}
\frac{d}{dt} \left(\frac{\partial V}{\partial\sigma}\right)_{\phi=\phi_0,\sigma=\sigma_0} = 0\ .
\eea
By applying the chain rule, $\dot{\sigma}_0$ can be expressed in terms of partial derivatives and $\dot{\phi}_0$ using the relation in Eq.~\eqref{eq:sigmadotpre}:
\bea\label{eq:sigmadotpre2}
\dot\sigma_0 = -\frac{\partial^2V}{\partial\sigma_0\partial\phi_0} \left(\frac{\partial^2 V}{\partial\sigma_0^2}\right)^{-1} \dot\phi_0 \ .
\eea
The partial derivatives in the above equation are evaluated at the meta-stable minimum $\sigma_0 = v_2$. It is evident that when $\partial^2 V/\partial\sigma_0^2 = 0$, the value $v_1$ must coincide with $v_2$, indicating that the phase transition becomes second-order. Consequently, the value of $\partial^2V/\partial\sigma^2$ must be proportional to $\Delta$, where $\Delta$ in defined in Eq.~(\ref{eq:Delta0}). 

Around the phase transition point, in terms of the potential defined in Eq.~(\ref{eq:Vsigma}), we find
\bea
\left.\frac{\partial^2 V(\phi_0,\sigma_0)}{\partial\sigma_0^2}\right|_{\sigma_0 = v_2} = 2 A v_1^4 \Delta(2 + \Delta) (1+\Delta)^2 \ .
\eea
Similarly, we can apply the same procedure used previously to calculate the cubic coupling for the curvature perturbation in order to determine the remaining partial derivatives in \eqref{eq:sigmadotpre2}:
\bea
\frac{\partial^2 V}{\partial\phi_0\partial\sigma_0} = -\frac{\eta_4 \kappa m_S^3}{M_{\rm pl}} \ ,
\eea
where $\eta_4$ is an order-one parameter. For the potential given in Eqs.~(\ref{eq:V}) and (\ref{eq:Vsigma}), we have
\bea
\frac{\partial^2 V}{\partial\phi_0\partial\sigma_0} &=& -2 c\phi_0 \sigma_0 = \frac{4\kappa}{M_{\rm pl}} \times \Big(-\frac{1}{2} c \phi_0^2 \sigma_0\Big) \ .
\eea

We observe that $\dot\sigma_0$ is proportional to $\dot\phi_0$, as the evolution $\phi_0$ induces the evolution of $\sigma_0$. As previously explained, $\dot\sigma_0$ is inversely proportional to $\Delta$. From Fig.~\ref{fig:S4}, we can see that, for a significant portion of the parameter space, $\Delta \ll 1$ when the phase transition begins to complete. Consequently, we expect an enhanced contribution to the non-Gaussianity. 

As a result, for generic models of the spectator sector with power-law interactions between the inflaton field and the spectator field, as shown in Eq.~(\ref{eq:V}), we have 
\bea\label{eq:sigmadot}
\dot\sigma_0 = \frac{\kappa \eta_5 m_S \dot \phi_0}{M_{\rm pl} \Delta} \ ,
\eea
where $\eta_5$ encapsulates all the ${\cal O}(1)$ effects. By substituting Eqs.~ (\ref{eq:eta1}), (\ref{eq:eta2}), (\ref{eq:eta3}), and (\ref{eq:sigmadot}) into Eq.~(\ref{eq:3pt}), we obtain the three-point interaction of the curvature perturbation: 
\bea\label{eq:R3}
\lambda(\tau) {\cal R}^3 \ ,
\eea
where 
\bea
\lambda(\tau) = \frac{\kappa^3 m_S^4 \dot\phi_0^3}{M_{\rm pl}^3 H^3} \left[\frac{\eta_1\eta_5}{2\Delta} \textcolor{black}{+}  \frac{\eta_2 \eta_5^2}{2\Delta^2} \textcolor{black}{-}\frac{\eta_3 \eta_5^3}{6\Delta^3} \right]  \ . 
\eea

In the symmetric phase, due to the ${\cal Z}_2$ symmetry, no ${\cal R}^3$ interaction is generated through the spectator sector, resulting in a vanishing contribution to the three-point non-Gaussianity. 
It is important to note that the three-point interaction can still be generated through loop diagrams. However, these contributions are inherently small and thus can be reasonably neglected.

\section{Calculation of the non-gaussianity}
\label{sec:NG}

We employ the standard method for calculating non-Gaussianity~\cite{Weinberg:2005vy,Weinberg:2006ac,Chen2017}. Specifically, the three-point correlation function for $\mathcal{R}$ in our model corresponds to a contact diagram with a time-dependent coupling. Thus, the reduced three-point correlation function of $\mathcal{R}$ can be expressed as follows:
\bea\label{eq:R3I}
\langle {\cal R}({\bf k}_1) {\cal R}({\bf k}_2) {\cal R}({\bf k}_3) \rangle' = \frac{3}{4}\int_{-\infty}^0 d\tau a^4(\tau) \lambda(\tau) \left[\frac{H^4}{\dot\phi_0^2}\right]^3  \frac{\tau^3}{k_1^2 k_2^2 k_3^2} f(k_1,k_2,k_3,\tau) \ ,
\eea
where $\lambda(\tau)$ represents the three-point coupling of the curvature perturbation $\mathcal{R}$, as given in Eq.~(\ref{eq:R3}). Note that the prime indicates that the delta function associated with momentum conservation has been omitted.
The remaining terms in equation \eqref{eq:R3I} originate from the inflaton bulk-to-boundary propagators. For convenience in subsequent calculations, we define the function $f(k_1,k_2,k_3,\tau)$ as follows:
\bea
f(k_1,k_2,k_3,\tau) =
{\rm Re} \left[ \left(1+\frac{i}{k_1\tau}\right) \left(1+\frac{i}{k_2\tau}\right) \left(1+\frac{i}{k_3\tau}\right) e^{i(k_1+k_2+k_3)\tau}\right]. 
\eea

Before performing explicit calculations, we can first analyze each term in equation \eqref{eq:R3I}. For the time-dependent coupling $\lambda(\tau)$, when $\tau$ is far from $\tau_*$ (the conformal time at the phase transition), the speed of its evolution is proportional to the slow-roll parameters. However, as $\tau$ approaches $\tau_*$, and in the parameter regime where $\Delta \ll 1$ at $\tau_*$, substituting into equation \eqref{eq:R3} yields:
\bea
\frac{\partial\lambda}{\partial\tau} \sim \frac{\partial\lambda}{\partial\Delta} \frac{\partial\Delta}{\partial\tau} \sim \frac{1}{\Delta^4(\tau_*)} \ .
\eea
Thus, for the integral over $\tau$ in Eq.~(\ref{eq:R3I}), we can perform a leading order estimation by assuming that all parameters in $\lambda$, except for $\Delta$, remain relatively constant with respect to $\tau$. This simplification allows us to focus on the enhancement induced by $\Delta$, which plays an important role in calculating $f_{\rm NL}$.

Before the modes (${\bf k}_1$, ${\bf k}_2$, ${\bf k}_3$ in (\ref{eq:R3I})) exit the horizon, the factor $f(k_1,k_2,k_3)$ oscillates rapidly with $\tau$, leading to a suppressed contribution to the $\tau$ integral. However, for modes that are completely outside the horizon, this factor can be approximated as a constant:
\bea
f(k_1,k_2,k_3) \rightarrow \frac{1}{3} \frac{k_1^3 + k_2^3 +k_3^3}{k_1 k_2 k_3}.
\eea
This observation indicates that for a leading-order estimation, the UV limit of the $\tau$ integral can be set at $\tau_{\rm UV} = -k_{\rm max}^{-1}$, where $k_{\rm max}$ denotes the largest value among $k_1$, $k_2$, and $k_3$. 

Considering the behavior of $\lambda(\tau)$ and $f(k_1, k_2, k_3)$ as discussed above, we can now highlight a crucial feature of the integral in Eq.~\eqref{eq:R3I}. Specifically, before $\tau$ reaches $\tau_*$, the $\tau$ integral takes the form $\int d\tau (1/\tau)$, indicating that each scale contributes equally to the integral until the phase transition is complete.

To qualitatively estimate $f_{\rm NL}$, we focus on the region that $\Delta \ll 1$ in this section. We neglect the variations of $\kappa$ and $\dot\phi_0$ in the $\tau$ integral of Eq.~(\ref{eq:R3I}). Furthermore, based on the specific interaction between the inflaton and spectator field in Eq.~\eqref{eq:V} and the general parameterization for a first-order phase transition potential in Eq.~\eqref{eq:Vsigma}, the parameters in the specific interaction can be re-expressed in terms of those in the general parameterization for convenience in subsequent calculations:
\bea\label{eq:Av1v2}
A v_1^2 v_2^2 = m_0^2 - c\phi_0^2 \ .
\eea
By substituting $v_0$ and $\Delta$ for $v_1$ and $v_2$ in \eqref{eq:Delta0}, and then taking the derivative with respect to $t$ on sides of the equation above, we find
\bea\label{eq:getc}
\frac{A v_0^4 \Delta\dot\Delta(1+\Delta)(2+\Delta)}{[1+(1+\Delta)^2]^3} = c \phi_0 \dot\phi_0 = - \kappa c\phi_0^2 \sqrt{2\epsilon} H \ .
\eea
Thus, assuming $A v_0^2$ and $c\phi_0^2$ are of the same order of magnitude, we derive the following expression in the regime of small $\Delta$: 
\bea
\frac{d}{dt} \Delta^2 = - \eta_6 \kappa H \epsilon^{1/2} \ ,
\eea
where $\eta_6$ is a parameter of order one. 
Given that we are in the slow-roll regime, where the slow-roll parameter and the Hubble parameter can be well approximated as constants, the above differential equation can be solved using the initial conditions at the time of the phase transition. We should reiterate that quantities evaluated at the phase transition are denoted by the subscript $*$. Therefore, we obtain:
\bea
\Delta^2 &\approx& \Delta_*^2 - \eta_6 \kappa H \epsilon^{1/2} (t - t_*)=\Delta_*^2 + \eta_6 \kappa \epsilon^{1/2} \log\left(\frac{\tau}{\tau_*}\right) \ .
\eea
As a result, the $\tau$ integral in Eq.~(\ref{eq:R3I}) gives the structure
\bea
&&\int_{\tau_{\rm UV}}^{\tau_*} \frac{d\tau}{\tau} \frac{1}{\Delta^n} \approx \frac{1}{\eta_6\kappa\epsilon^{1/2}}\int_{\Delta_{\rm UV}^2}^{\Delta_*^2} \frac{d\Delta^2}{\Delta^n}= \left\{ \begin{array}{ll} \frac{2}{\eta_6\kappa \epsilon^{1/2}\textcolor{black}{(2-n)}} \left( \frac{1}{\Delta_*^{n-2}} - \frac{1}{\Delta_{\rm UV}^{n-2}} \right) \ , ~&~ n \neq 2 
 \\
 \frac{2}{\eta_6 \kappa \epsilon^{1/2}} \log\left( \frac{\Delta_*}{\Delta_{\rm UV}} \right) \ , ~&~ n = 2 \\
 \end{array}
 \right.
\eea
where $\Delta_{\rm UV}^2 = \Delta_*^2 - \eta_6 \kappa \epsilon^{1/2} \log(-k_{\rm max}\tau_*) = \Delta_*^2 + \eta_6 \kappa \epsilon^{1/2} N_{\rm max}$. Here, 
\bea\label{eq:defofNmax}
N_{\rm max} \equiv |\log(-k_{\rm max}\tau_*)|\ , 
\eea
represents the number of e-folds from the exit of the hardest mode from the horizon until the phase transition. 

Then, the bi-spectrum (\ref{eq:R3I}) can be expressed 
as
\bea
B({\bf k_1},{\bf k_2},{\bf k_3})&\approx&\frac{m_S^4 \kappa^2}{2 M_{\rm pl}^2 H^2 \eta_6}  \left(\frac{H^4}{\dot\phi_0^2}\right)^2  \frac{k_1^3 + k_2^3 + k_3^3}{k_1^3 k_2^3 k_3^3}\nnn
&&\times  \left[  -\frac{\eta_1\eta_5}{2}g(1,N_{\rm max}) - \frac{\eta_2 \eta_5^2}{2}g(2,N_{\rm max}) + \frac{\eta_3 \eta_5^3}{6}g(3,N_{\rm max})\right] \ . 
\eea
The form factor $g(n,N)$ is given by
\bea
g(n,N) = \left\{ \begin{array}{ll}  \frac{1}{2-n}\left(\frac{1}{\Delta_*^{n-2}} - \frac{1}{(\Delta_*^2+\eta_6\kappa\epsilon^{1/2} N_{\rm max})^{n/2-1}}\right)  \ , ~&~ n \neq 2 
 \\
  \log\left[ \frac{\Delta_*}{(\Delta_*^2+\eta_6\kappa\epsilon^{1/2} N_{\rm max})^{1/2}} \right] \ , ~&~ n = 2 \\
 \end{array}
 \right.
\eea
In the squeezed limit, where $k_{\rm max}\approx k_1 \approx k_2 \gg k_3$, we find a scaling non-gaussianity:
\bea
f_{\rm NL}(N_{\rm max})
&=& \frac{m_S^4 \kappa^2}{M_{\rm pl}^2 H^2 \eta_6} \left[-  \eta_1\eta_5g(1,N_{\rm max}) -\eta_2 \eta_5^2g(2,N_{\rm max}) + \frac{\eta_3 \eta_5^3}{3}g(3,N_{\rm max})\right] \ .
\eea
It is evident that through the dependence on $N_{\rm max}$, $f_{\rm NL}$ is a function of $k_{\rm max}$. 
From the Friedmann equation, we have $M_{\rm pl}^2 H^2 \sim \rho_{\rm inf}$, where $\rho_{\rm inf}$ is the total energy density of the universe during inflation. In the expression for $f_{\rm NL}$, the factor $m_S^4$ is the fourth-power of the characteristic energy scale of the spectator sector; therefore, we can use $m_S^4$ to estimate the latent energy density released during the phase transition. Consequently, we can perform the following estimation:
\bea
\frac{m_S^4}{M_{\rm pl}^2 H^2} = \eta_8\times \frac{L}{\rho_{\rm inf}} \ ,
\eea
where $L$ is the latent energy density released during the phase transition and $\rho_{\rm inf}$ is the total energy density of the universe during inflation. Thus, $f_{\rm NL}$ can be further simplified to 
\bea\label{eq:fNL1}
f_{\rm NL}(N_{\rm max})
&=& \frac{\eta_8 \kappa^2}{\eta_6}\frac{L}{\rho_{\rm inf}} \left[ -\eta_1\eta_5 g(1,N_{\rm max})- \eta_2 \eta_5^2 g(2,N_{\rm max})+ \frac{\eta_3 \eta_5^3}{3}g(3,N_{\rm max})\right] \ .
\eea
It is important to note that the dimensionless parameters $\eta_1$, $\eta_2$, $\eta_3$, $\eta_5$, and $\eta_6$ are $\mathcal{O}(1)$ factors. The parameter $\eta_8$ represents the ratio of latent energy density to inflationary energy density, which must be less than 1. Consequently, $f_{\text{NL}}$ is primarily governed by $\Delta^2_*$ and $N_{\text{max}}$. 

\begin{figure}[t]
 \centering
   \includegraphics[width=0.7\textwidth]{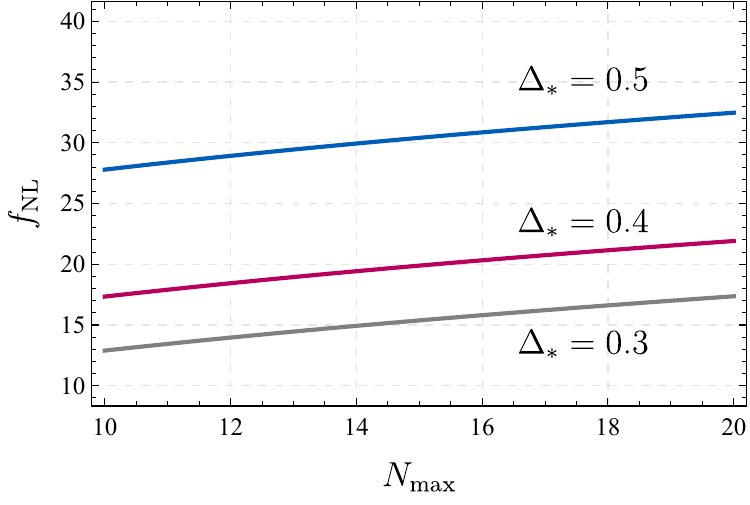}
\caption{The dependence of  $f_{\text{NL}}$  on $ N_{\text{max}}$  is illustrated, with parameter ratios  $\eta_1 = \eta_2 = \eta_3 = \eta_5 = \eta_6 = \eta_7 = 1$ ,  $\eta_8 = 0.5$ , and  $\kappa = 20$. }
\label{fig:fNL1}
\end{figure}

In Fig. \ref{fig:fNL1}, we present $f_{\text{NL}}$ as a function of $N_{\text{max}}$ for several representative values of $\Delta^2_*$.
From the figure, it is evident that the $f_{\rm NL}$ induced by the phase transition is proportional to $L$, the latent energy density. Furthermore, it exhibits a slight dependence on $k_{\rm max}$ through $N_{\rm max}$. This dependency generates a scaling behavior in $f_{\rm NL}$ and may serve as a notable feature of the non-Gaussian signals produced by phase transitions during inflation. 
To quantify this mild scale dependence, we adopt the conventional definition from previous literature~\cite{Chen:2005fe, Kumar:2009ge, Byrnes:2009pe, Planck:2019kim}, where the running index of $f_{\rm NL}$ is defined as
\bea
n_{\rm NG} = \frac{\partial \ln f_{\rm NL}}{\partial \ln k_{\rm max}} \ .
\eea
Then, from Eq.~(\ref{eq:fNL1}), we obtain 
\bea
n_{\rm NG} \approx \frac{-\eta_1 g'(1,N_{\rm max}) - \eta_2 \eta_5 g'(2,N_{\rm max})+ \frac{1}{3} \eta_5^2 g'(3,N_{\rm max})}{-\eta_1 g(1,N_{\rm max}) - \eta_2 \eta_5 g(2,N_{\rm max}) + \frac{1}{3} \eta_5^2 g(3,N_{\rm max})} \ ,
\eea
where the primes on $g(n,N_{\rm max})$ indicate the derivative with respect to $N_{\rm max}$.
In the regime where $\Delta \ll 1$ at the completion of the phase transition, the $g(3,N_{\rm max})$ term, specifically the third term in the square bracket of Eq.~(\ref{eq:fNL1}), contributes most significantly to $f_{\rm NL}$ due to its proportionality to $\Delta^{-3}$. 

%
%
Thus, as a good approximation, for qualitative analysis, we can keep only the $g(3,N_{\rm max})$ contribution, and in this case $n_{\rm NG}$ can be approximately written as 
\bea
n_{\rm NG} \approx \frac{g'(3,N_{\rm max})}{g(3,N_{\rm max})} \ .
\eea
In the regime where $\Delta_{\rm UV} \gg \Delta_*$, the expression for $n_{\rm NG}$ can be further simplified to 
\bea
n_{\rm NG} \approx \frac{\Delta_*}{(\eta_6 \kappa \epsilon^{1/2})^{1/2} N_{\rm max}^{3/2}} \ .
\eea
By selecting values for $\kappa$ and $\eta_6$ on the order of one, with $\epsilon \sim {\cal O}(10^{-5})$,
$N_{\rm max} \sim {\cal O}(10)$, and $\Delta_*\sim 0.1$, we find that $n_{\rm NG}\sim {\cal O}(0.1)$. 

\begin{figure}
 \centering
   \includegraphics[width=0.7\textwidth]{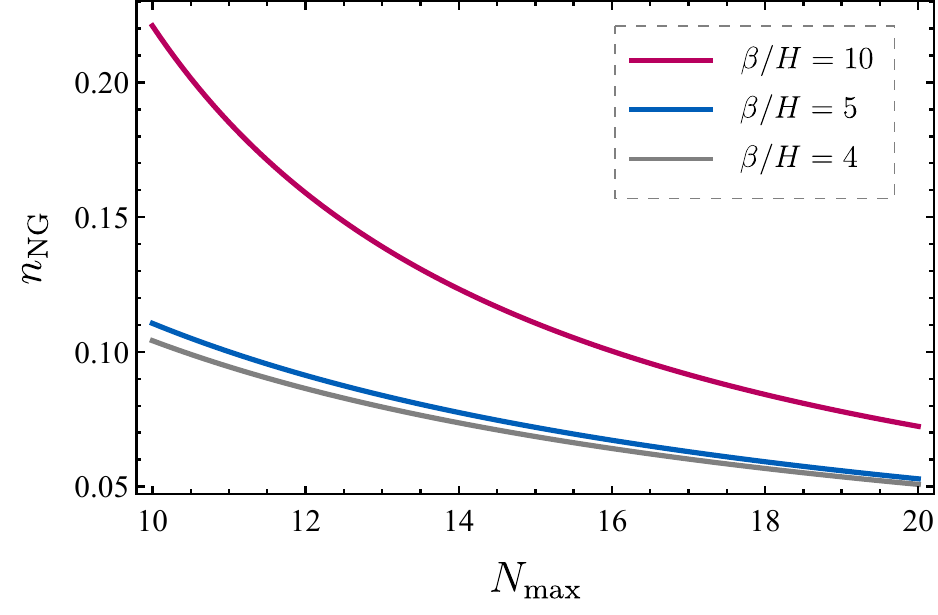}
\caption{The running index $n_{\rm NG}$ as a function of $N_\text{max}$ with $\kappa = 1$, $\theta  = 1/10$ and $\epsilon =  10^{-4}$.}
\label{fig:nng}
\end{figure}

\section{Non-Gaussianities vs GWs}
\label{sec:NGvsGW}

Both the primary and secondary GW signals induced by first-order phase transitions during inflation have been studied in the literature~\cite{An:2020fff,An:2022cce,An:2022toi,An:2023idh,An:2023jxf}. The frequencies of the GWs are significantly influenced by the specific e-fold in which the phase transition occurs. Notably, the shapes--- particularly the infrared (IR) region---of the GW spectrum for both the primary and secondary GWs remain independent of the specifics of the phase transition models. The peak values of the GWs are determined by the parameters $(H/\beta)$, $\theta \equiv L/\rho_{\rm inf}$, and $\kappa$. Here, $\beta$ is defined as 
\bea
\beta = - \frac{d {\cal S}_E}{d t} \ . 
\eea
In this section, we establish the relation between the GW signal and $f_{\rm NL}$ examined in the previous sections. We will not restrict our analysis to the assumption that $\Delta_* \ll 1$ at the completion of the phase transition. 

To develop the relationship between the GW signal and $f_{\rm NL}$, we first replace $c$ with the phenomenological parameter $\beta$. 
From Eq.~(\ref{eq:getc}), we can get 
\bea\label{eq:getc2}
c = \frac{A v_0^4 \Delta_* \dot\Delta_* (1+\Delta_*)(2+\Delta_*)}{\phi_{0*}\dot\phi_{0*}[2+\Delta_*(2+\Delta_*)]^3} \ ,
\eea
where the asterisk ($*$) indicates the values of the corresponding quantities are evaluated at the time $t_*$. 
In Eq.~(\ref{eq:getc2}), $\Delta_*$ can be estimated from the bounce action ${\cal S}_E$ using a given value of $A v_0^2$. The value for $v_0^2$ can be calculated from the latent energy density $L$. Consequently, the coupling $c$ can be determined using $A v_0^2$, $L$, $\Delta_*$, $\dot\Delta_*$, $\phi_{0*}$, and $\dot\phi_{0*}$. Furthermore, the bounce action is described by Eq.~(\ref{eq:SE1}), leading to the following expression: 
\bea\label{eq:defofbetas}
\beta_* = - \left.\frac{d {\cal S}_E}{d t}\right|_{t_*} = - \frac{\hat S'(\Delta_*) \dot\Delta_*}{A v_0^2} \ .
\eea
Thus, the $\phi-\sigma$ coupling $c$ can be further simplified as: 
\bea\label{eq:c0}
c =  \frac{A^2 v_0^6 \kappa_*}{M_{\rm pl}^2 \sqrt{2\epsilon_*}} \frac{\beta_*}{H}\frac{\Delta_*(1+\Delta_*)(2+\Delta_*)}{[1+(1+\Delta_*)^2]^3 \hat S'(\Delta_*)} \ .
\eea
As expected, $c$ is suppressed by $M_{\rm pl}^2$, since the direct interaction of $\phi$ with the spectator sector violates the flatness of the inflaton potential. With the knowledge of $c$, we can calculate $m_0^2$ using the following equation: 
\bea\label{eq:m0}
A v_0^4 \frac{(1+\Delta_*)^2}{[1+(1+\Delta_*)^2]^2} = m_0^2 - c \phi_{0*}^2 \ .
\eea
We can further compute
\bea\label{eq:sigmadot2}
\dot\sigma_0 = \frac{v_0 \dot\Delta}{[1+(1+\Delta)^2]^{3/2}} \ ,
\eea
where $\dot\Delta$ can be derived from Eq.~(\ref{eq:getc}) using $\Delta$, $\phi_0$ and $\dot\phi_0$. We finally arrive at an expression of $\dot\sigma_0$: 
\bea\label{eq:sigmadot3}
\dot\sigma_0 = - \frac{A v_0^3 \phi_0 \dot\phi_0 \kappa_*}{M_{\rm pl}^2 (2\epsilon_*)^{1/2}} \frac{\beta_*}{H} f_1(\Delta,\Delta_*) \ ,
\eea
where
\bea
f_1(\Delta,\Delta_*) = \frac{[1+(1+\Delta)^2]^{3/2}}{\Delta(1+\Delta)(2+\Delta)} \frac{\Delta_*(1+\Delta_*)(2+\Delta_*)}{[1+(1+\Delta_*)^2]^3 \hat S'(\Delta_*)} \ .
\eea
Furthermore, $\Delta$ can be obtained by solving Eq.~(\ref{eq:Av1v2}), yielding
\bea\label{eq:Delta}
\Delta = \frac{1}{2} \left( w^{1/2} + \sqrt{w-4} \right) -1 \ ,
\eea
where 
\bea\label{eq:w}
w = \frac{A v_0^4}{ m_0^2 - c\phi_0^2} \ .
\eea
\begin{figure}[t]
 \centering
   \includegraphics[width=0.49\textwidth]{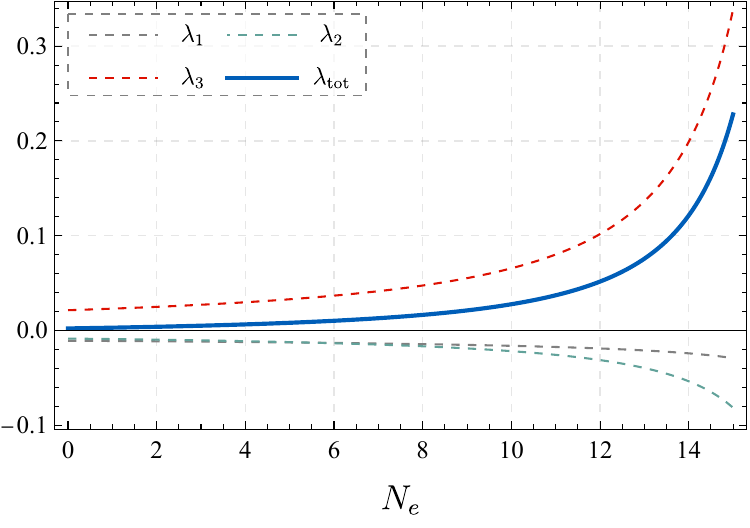}
   \includegraphics[width=0.49\textwidth]{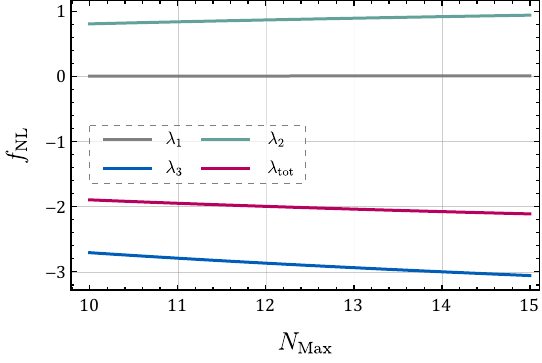}
\caption{Left panel: Couplings as a function of the e-folding number. Right panel: $f_{\text{NL}}$ as a function of $N_{\text{max}}$. The different colors represent the contributions of different couplings. In both panels, the following parameters are used: $\epsilon=10^{-5}$, $\theta=1/200$, $\beta/H=20$, $\kappa=1$.}
\label{fig:couple}
\end{figure}

Using Eqs.~(\ref{eq:c0}), (\ref{eq:m0}), and (\ref{eq:sigmadot3}), we can express the three-point coupling of the curvature perturbation $\lambda$ in Eq.~(\ref{eq:3pt}) as a function of $\phi_0$, $\dot\phi_0$, $\phi_{0*}$, $\dot\phi_{0*}$, $\beta/H$, $A v_0^2$, $\Delta_*$, $S_{E*}$, and $L$. The first four variables are related to the inflaton sector, while the last four are related to the phase transition sector, which are strongly connected to the GW signal. The dependence on $\Delta$ and $\Delta_*$ from each of the three terms in Eq.~(\ref{eq:3pt}) are factorized. After some tedious yet straightforward calculations, we obtain
\bea
\lambda_1 &=& \left(\frac{\beta_*}{H}\right)^2 \frac{L\dot\phi_0^3 \phi_0\kappa_*^2\Delta_*^2}{H^3 M_{\rm pl}^4{\cal S}_{E*}^2 \epsilon_*} g_1(\Delta) h_1(\Delta_*) \nnn
\lambda_2 &=& \left(\frac{\beta_*}{H}\right)^3 \frac{L\dot\phi_0^3 \phi_0^3\kappa_*^3\Delta_*^3}{H^3 M_{\rm pl}^6{\cal S}_{E*}^3 \epsilon_*^{3/2}} g_2(\Delta) h_2(\Delta_*) \nnn
\lambda_3 &=& \left(\frac{\beta_*}{H}\right)^3 \frac{L\dot\phi_0^3 \phi_0^3\kappa_*^3\Delta_*^3}{H^3 M_{\rm pl}^6{\cal S}_{E*}^3 \epsilon_*^{3/2}} g_3(\Delta) h_3(\Delta_*) \nnn
\eea
We have 
\bea\label{eq:gandh}
g_1(\Delta) &=& \frac{1+(1+\Delta)^2}{\Delta(2+\Delta)} \nnn
h_1(\Delta_*) &=&  \frac{12 (2+\Delta_*)^2}{(1+\Delta_*)^2(-2+\Delta_*(2+\Delta_*))(2+\Delta_*(2+\Delta_*))^3} \left(\frac{\hat S(\Delta_*)}{\hat S'(\Delta_*)}\right)^2 \nnn
g_2(\Delta) &=& \frac{(2 + \Delta (2 + \Delta))^3}{(\Delta^2 (1 + \Delta)^2 (2 + \Delta)^2)} \nnn
h_2(\Delta_*) &=& -\frac{6\sqrt{2}  (2+\Delta_*)^3 }{(2+\Delta_*(2+\Delta_*))^6 (2 - \Delta_*^2 (3+\Delta_*))} \left(\frac{\hat S(\Delta_*)}{\hat S'(\Delta_*)}\right)^3 \nnn
g_3(\Delta) &=& \frac{(2+\Delta(2+\Delta))^3 (4+7\Delta(2+\Delta))}{\Delta^3(1+\Delta)^2 (2+\Delta)^3} \nnn
h_3(\Delta_*) &=& \frac{\sqrt{2} (2+\Delta_*)^3}{(2+\Delta_*(2+\Delta_*))^6 (2-\Delta_*^2(3+\Delta_*))}\left(\frac{\hat S(\Delta_*)}{\hat S'(\Delta_*)}\right)^3 \ .
\eea

Before presenting the numerical results, we first examine the qualitative magnitude of $f_{\rm NL}$. From equation (\ref{eq:gandh}), it is clear that the magnitude of $\lambda_3$ consistently exceeds that of $\lambda_2$. In Fig. \ref{fig:couple}, we show the running of the coupling constants for each component as a function of the e-folding number, with the parameter choices specified in the caption. As demonstrated in Fig. \ref{fig:couple}, $\lambda_3$ indeed dominates the contribution to $f_{\text{NL}}$.

During the slow roll inflation, we first neglect the change of $\phi_0$ and $\dot\phi_0$. We also assume $H$ remains constant. Under these assumptions, the $\tau$ integral in (\ref{eq:R3I}) is proportional to 
\bea\label{eq:fnlgeneral}
\int \frac{d\tau}{\tau} f(k_1,k_2,k_3,\tau) {\cal G}(\Delta(\tau)) \ ,
\eea
where ${\cal G}$ collects all the $\Delta$ dependence of the integrand. As discussed in Sec.~\ref{sec:NG}, in the region that $|k_{\rm max} \tau| > 1$, $f(k_1,k_2,k_3,\tau)$ oscillates with $\tau$, making its contribution to the $\tau$ integral less significant compared to the region where $|k_{\rm max}\tau| < 1$, in which $f(k_1,k_2,k_3,\tau)$ becomes approximately constant. 

\begin{figure}[t]
 \centering
   \includegraphics[width=0.49\textwidth]{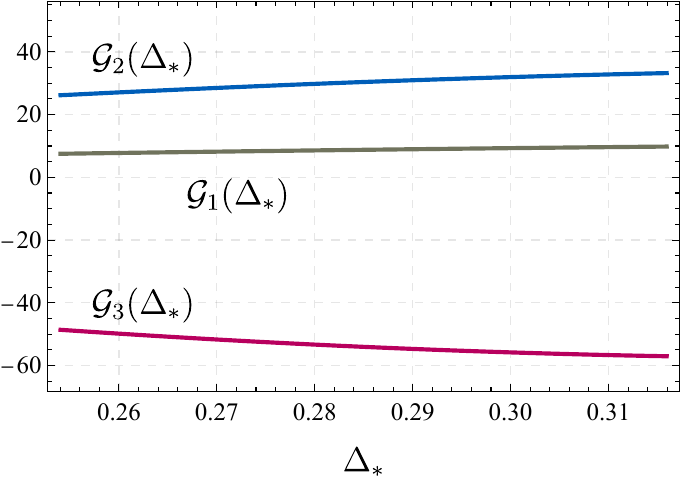}
   \includegraphics[width=0.49\textwidth]{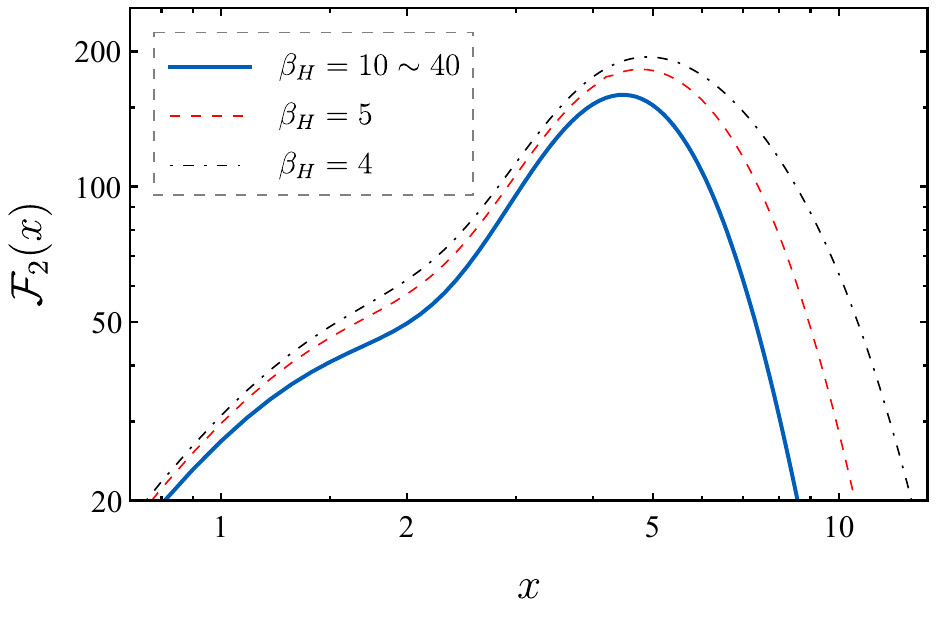}
\caption{Left panel: $\ma{G}_{1,2,3}$ as functions of $\Delta_*$ with $\epsilon_* = 10^{-5}$. Right panel: The profile of the GW spectrum $\ma{F}_2$ with different parameter choice.}
\label{fig:fnlandF2}
\end{figure}

To provide an estimate of the $f_{\text{NL}}$ induced by the phase transition, it is necessary to express $f_{\text{NL}}$ in terms of relevant physical parameters, such as the latent heat $L$, among others. The total magnitude of the non-Gaussianity parameter, $f_{\text{NL}}$, arises from three distinct contributing components. By substituting $\lambda_1$, $\lambda_2$ and $\lambda_3$ into equation \eqref{eq:fnlgeneral} and completing the $\tau$ integral within the region where $|k_{\rm max} \tau| < 1$, we obtain the expression for the bispectrum:

\begin{align}
B({\bf k_1},{\bf k_2},{\bf k_3}) \approx  \fr{1}{4} \fr{k_1^3+k_2^3+k_3^3}{k_1^3k_2^3k_3^3} & \Bigg[ \left(\fr{H^4}{\dot{\phi}_0^2}\right)^2 \left( \fr{\beta_*}{H} \right)^2 \left( \fr{L}{\rho_\text{inf}} \right) \left( \fr{\Delta_*}{S_{E*}} \right)^2 \int \fr{\d \tau }{(\epsilon_*)^{1/2} \tau} g_1(\Delta) h_1(\Delta_*) \nonumber \\
&  + \left(\fr{H^4}{\dot{\phi}_0^2}\right)^2 \left( \fr{\beta_*}{H} \right)^3 \left( \fr{L}{\rho_\text{inf}} \right) \left( \fr{\Delta_*}{S_{E*}} \right)^3 \int \fr{\d \tau }{\epsilon_* \tau} g_2(\Delta) h_2(\Delta_*) \nonumber \\
& + \left(\fr{H^4}{\dot{\phi}_0^2}\right)^2 \left( \fr{\beta_*}{H} \right)^3 \left( \fr{L}{\rho_\text{inf}} \right) \left( \fr{\Delta_*}{S_{E*}} \right)^3 \int \fr{\d \tau }{\epsilon_* \tau} g_3(\Delta) h_3(\Delta_*) \Bigg] \label{eq:bispectrumgeneral}
\end{align}
Specifically, only $\mathcal{G}_1$ exhibits an additional dependence on the slow-roll parameter $\epsilon$, while the others only show dependence on $\Delta_*$. This arises because the integral is taken over $\tau$, whereas the integrand is expressed in terms of $\Delta$. 
With this in mind, we can now explicitly present the expression for the amplitude of non-Gaussianity, $f_{\text{NL}}$, as derived from equation \eqref{eq:bispectrumgeneral}:


\begin{equation}\label{eq:fnlgeneral1}
\begin{split}
    f_{\text{NL}}&=\Big(\frac{\beta_*}{H}\Big)^2\Big(\frac{L}{\rho_{\text{inf}}}\Big)\Big(\frac{\Delta_{*}}{\ma{S}_{E*}}\Big)^2\Big(\frac{M_{\rm pl}}{\phi}\Big)\ma{G}_1(\Delta_*)\\
    &~~~+\Big(\frac{\beta_*}{H}\Big)^3\Big(\frac{L}{\rho_{\text{inf}}}\Big)\Big(\frac{\Delta_{*}}{\ma{S}_{E*}}\Big)^3\ma{G}_2(\Delta_*)\\
    &~~~+\Big(\frac{\beta_*}{H}\Big)^3\Big(\frac{L}{\rho_{\text{inf}}}\Big)\Big(\frac{\Delta_{*}}{\ma{S}_{E*}}\Big)^3\ma{G}_3(\Delta_*),
\end{split}
\end{equation}
where 
\begin{align}
\ma{G}_1(\Delta_*) =  \int \fr{\d \tau }{(\epsilon_*)^{1/2} \tau} g_1(\Delta) h_1(\Delta_*), \quad \ma{G}_{2,3}(\Delta_*) =  \int \fr{\d \tau }{\epsilon_* \tau} g_{2,3}(\Delta) h_{2,3}(\Delta_*) \ ,
\end{align}
where we introduce the functions $\mathcal{G}_{1,2,3}(\Delta_*)$ to encapsulate the integral in Eq.~\eqref{eq:fnlgeneral}, thereby simplifying our notation. We present the numerical result for  $\mathcal{G}_{1,2,3}$ in Fig. \ref{fig:fnlandF2}, 
which display a mild dependence on $\Delta_*$. This integral, being highly model-dependent, plays a secondary role in illustrating the general features of non-Gaussianity arising from a first-order phase transition. The $f_{\text{NL}}$ associated with such a transition is predominantly governed by a few key physical parameters. In particular, a larger energy release during the phase transition leads to a correspondingly higher $f_{\text{NL}}$. Furthermore, as we will discuss, the non-Gaussianity signal complements gravitational wave signatures in relation to the parameter $\beta_*/H$.


\begin{figure}
 \centering
   \includegraphics[width=0.7\textwidth]{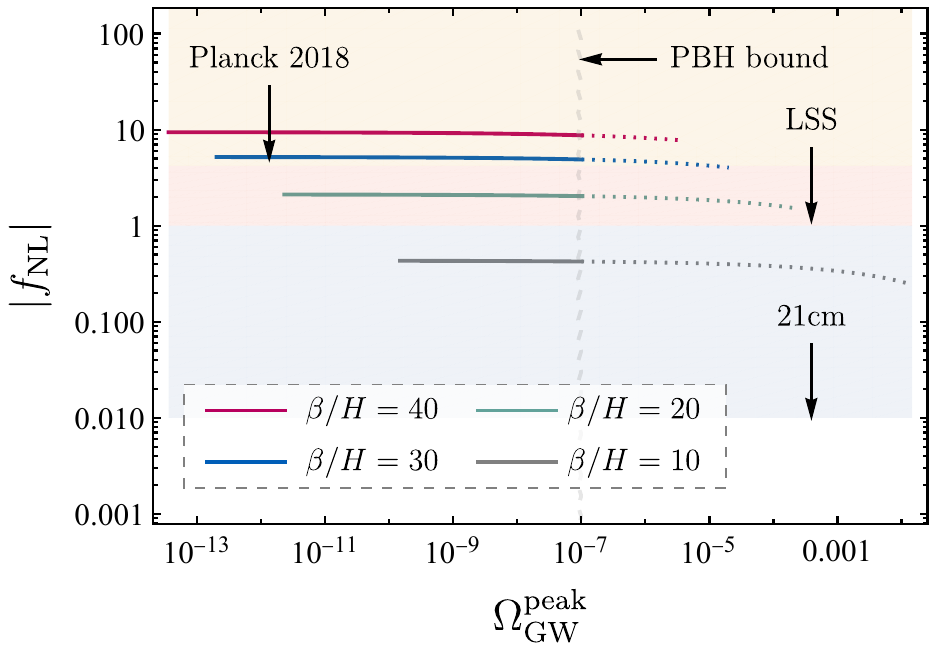}
\caption{The numerical results for the non-Gaussianity $f_\text{NL}$ as a function of the peak gravitational wave energy density $\Omega_\text{GW}^{\text{peak}}$ for the $\sigma^6$ model, evaluated over the range $\kappa_* \in (0.01, 1)$, with fixed parameters $\theta = 1/50$ and $\epsilon_* = 10^{-5}$. The yellow shaded region represents the sensitivity limits from Planck 2018~\cite{Planck:2019kim}, while the red and blue shaded regions denote the projected sensitivity of forthcoming Large-Scale Structure~\cite{SPHEREx:2014bgr} and 21 cm surveys~\cite{Munoz:2015eqa,Meerburg:2016zdz}. }
\label{fig:fNLvsGW}
\end{figure}

\begin{figure}
 \centering
   \includegraphics[width=0.7\textwidth]{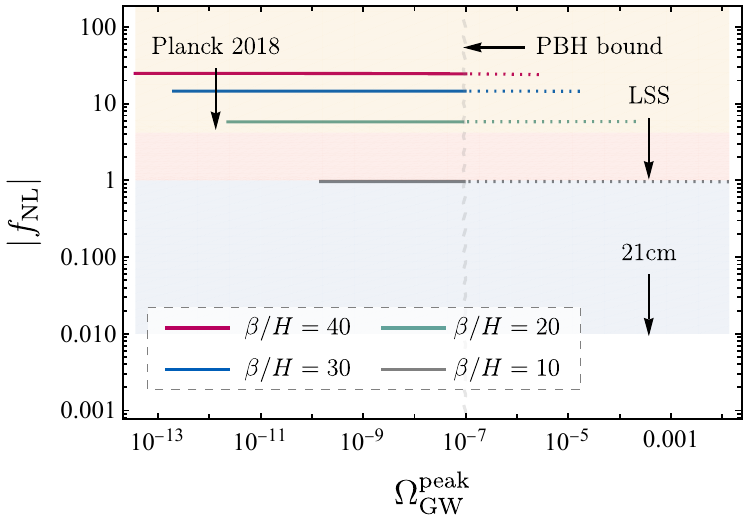}
\caption{The numerical results depict the  $ f_\text{NL}$  as a function of  $\Omega_\text{GW}^{\text{peak}}$  over the range  $\kappa_* \in (0.01, 1)$  for the CW model. The potential parameters are set to be $\lambda = 1$  and  $\rho = 1$ , while all other parameters match those used in Fig.~\ref{fig:fNLvsGW}. Shaded regions indicate the sensitivity limits, defined consistently with those in Fig.~\ref{fig:fNLvsGW}.}
\label{fig:fNLvsGWCW}
\end{figure}

\begin{figure}
 \centering
   \includegraphics[width=0.7\textwidth]{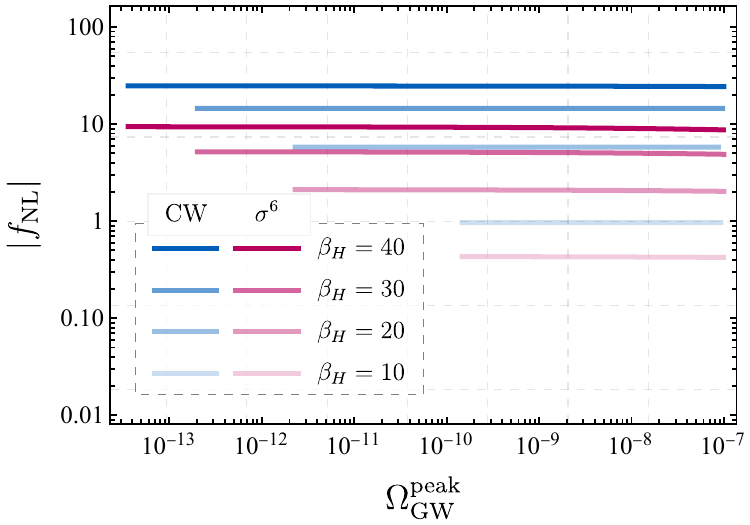}
\caption{Comparison of the numerical results for  $f_\text{NL}$  versus  $\Omega_\text{GW}^{\text{peak}}$  from the  $\sigma^6$  model and the CW model, with parameters set identically to those in Fig.~\ref{fig:fNLvsGW} and Fig.~\ref{fig:fNLvsGWCW}.}
\label{fig:CWvssigma6}
\end{figure}

Both the primary and secondary GW signals induced by first-order phase transitions during inflation have been studied~\cite{An:2020fff,An:2022cce,An:2022toi,An:2023idh,An:2023jxf}. The signal strength of the GW signals depend on various powers of $H/\beta$ and $L/\rho_{\rm inf}$, as well as $\kappa$. Specifically, the spectrum of the first and secondary GW are given by 
\begin{align}
\Omega_\text{GW}^{(1)}(f) = & \Omega_R \left( \fr{H}{\beta} \right)^5 \left( \fr{L}{\rho_\text{inf}} \right)^2 \ma{F}_1 (\fr{f}{f_\text{peak}}) \\
\Omega_\text{GW}^{(2)}(f) = & \fr{\Omega_R}{\epsilon^2} \left( \fr{H}{\beta} \right)^6 \left( \fr{L}{\rho_\text{inf}} \right)^4 \ma{F}_2 (\fr{f}{f_\text{peak}}) 
\end{align}
where $\Omega_\text{R}$ represents the radiation energy density of the universe, while $\epsilon$ denotes the slow-roll parameter following the phase transition. The functions $\ma{F}_{1,2}$ correspond to the shape profiles of the primary and secondary gravitational waves, with their detailed derivation provided in ~\cite{An:2020fff,An:2022cce,An:2022toi,An:2023idh,An:2023jxf}.  Notably, in comparison to the primary gravitational waves, the amplitude of the secondary gravitational waves generated during the phase transition is inherently amplified by the slow-roll parameter, potentially leading to a detectable signal in PTA, and therefore it will be the focus of our following discussions. Fig. \ref{fig:fnlandF2} illustrates the profile of $\ma{F}_2$, which exhibits a global peak. This peak will be utilized as a defining characteristic of the gravitational wave signal’s strength in subsequent analyses.

It is important to emphasize that the same set of parameters ($\epsilon$, $\beta$, and $\theta$) govern both the size of the curvature non-Gaussianity, $f_{\rm NL}$ and the GW spectrum. In Fig. \ref{fig:fNLvsGW}, we present the numerical results of $f_\text{NL}$ vs the peak amplitude of the secondary gravitational wave spectrum, $\Omega_\text{GW}^{\text{peak}}$ with different values of $\beta/H$. In the figure, the constraint on $f_{\rm NL}$ from the Planck result~\cite{Planck:2019kim} is also shown together with the future sensitivities of the Large Scale Structure Survey~\cite{SPHEREx:2014bgr} and 21cm survey~\cite{Munoz:2015eqa,Meerburg:2016zdz}. We can see that for a large part of the parameter space, the $f_{\rm NL}$ corresponding to the phase transition during inflation can be observed by future experiments.

\section{The Coleman-Weinberg potential case}
\label{sec:CW}

In the previous sections, our analysis focuses on the case where the field $\sigma$ is governed by a polynomial potential, a form commonly arising in the context of effective field theory. In this section, we shift focus to a different class of potentials, specifically the CW potential, which typically emerges from quantum corrections at the loop level. The analysis of the primordial non-Gaussianity, $f_{\rm NL}$ and the GW signal are in parallel to the polynomial potential case. Therefore, we only list the results in this section. 
To be specific, we consider the following potential for the spectator field $\sigma$,
\be\label{eq:VCW}
V_\sigma^{(\text{CW})} = \fr{1}{2} \mu_\text{eff}^2 \sigma^2 - \fr{\lambda}{4} \sigma^4 +\fr{\rho}{4} \sigma^4 \log \left(\fr{\sigma^2}{\Lambda^2}\right).
\ee
Here, as in the previous case, $\mu_\text{eff}^2 = - (m_\sigma^2 - c \phi_0^2)$ represents the running mass generated by the background of the rolling inflaton. As in the previous discussion, the evolution of $\mu_\text{eff}$ will ultimately induce a first-order phase transition in the spectator sector, producing a distinctive gravitational wave signal. A detailed analysis of this gravitational wave signal can be found in~\cite{An:2022cce}. The calculation of $f_{\rm NL}$ can be done in parallel to the case of the polynomial potential as discussed in previous sections. However, due to the logarithmic term, there are no analytical expressions for $c$ and $\lambda$ unlike in the case of the polynomial potential. Therefore, we only present the numerical results here. In Fig.~\ref{fig:fNLvsGWCW}, we present the plot of $|f_{\rm NL}|$ vs $\Omega_{\rm GW}^{\rm peak}$ for $\beta/H = 10, 20, 30, 40$, $\kappa_* \in (0.01,1)$, $\theta = 1/50$, and $\epsilon_* = 10^{-5}$. We can observe that in the CW potential case, just like in the case of the polynomial potential, in a large part of the parameter space, observable $f_{\rm NL}$ can appear corresponding to the GW signal. In Fig.~\ref{fig:CWvssigma6}, we show the comparison of the results of $f_{\rm NL}$ and the GW signal between the polynomial model and the CW model. We can see that for the same choices of the phenomenological parameters, such as $\beta/H$ and $\kappa$, $f_{\rm NL}$ in the two models only differ by an order one parameter. Therefore, we can believe that our conclusion that the phase transition in the spectator sector during inflation induces observable $f_{\rm NL}$ is robust against the choices of phase transition models.

\section{Summary and discussion}
\label{sec:summary}

In this work, we explore the non-Gaussianity generated alongside a phase transition that occurs during inflation. We assume that the phase transition takes place within a spectator sector and is triggered by the evolution of the inflaton field. Our findings indicate that, in general, the magnitude of the non-Gaussianity, denoted as $f_{\rm NL}$, is of order one and could be detectable in future cosmological observations. We derive that $f_{\rm NL}$ is proportional to $(\beta/H)^3$, while the gravitational wave signal, $\Omega_{\rm GW}$ is proportional to $(\beta/H)^{-6}$. Consequently, the primordial non-Gaussianity serves as a complement to the gravitational wave signal when searching for phase transitions during inflation. Moreover, as we demonstrate in Sections~\ref{sec:NGvsGW} and \ref{sec:CW}, there exists a substantial region of parameter space where both gravitational wave signals and $f_{\rm NL}$ can be observed in the future.
Furthermore, as discussed in Sec.~\ref{sec:NG}, since the $\tau$ integral in the calculation of $f_{\rm NL}$ sensitive to both the UV and IR cut-offs, the non-Gaussianity, $f_{\rm NL}$ produced in this way slightly depends on $k_{\rm max}$. The typical value of index $n_{\rm NG} = \partial\ln f_{\rm NL}/\partial\ln k_{\rm max}$ is about ${\cal O}(0.1)$. 

In this work, we focused on the scenario that a ${\cal Z}_2$ symmetry restoration first-order phase transition triggered by the evolution of the inflaton field during inflation as an illustration of the calculation for the non-Gaussianity corresponding to the phase transition. For symmetry-breaking phase transitions, in the symmetry-breaking phase, a 3pt interaction of the ${\cal R}$ will induced by the evolution of $\sigma$. Thus, in this case, the UV cutoff of the $\tau$ integral in Eq.~(\ref{eq:R3I}) is the time of the phase transition $\tau_\star$, instead of $k_{\rm max}^{-1}$. The IR cutoff can be set at the time when inflation ends. The form of the integrand is similar to the symmetry-restoration case. Thus, we expect that $f_{\rm NL}$ contributed by the spectator sector in the symmetry breaking case has the similar size of that from the symmetry restoration case as discussed in detail in this work. However, $f_{NL}$ produced in the symmetry breaking case will have no scale dependence. 

For second-order phase transitions, we also expect to have an ${\cal O}$(1) $f_{\rm NL}$. However, as we saw in the discussion in Sec.~\ref{sec:NG}, the $\tau$ integral around the phase transition point significantly contributes to $f_{\rm NL}$. The behavior of the spectator field at the phase transition point differs, necessitating a distinct approach for second-order phase transitions. We leave the discussion for second-order phase transition in future work.

\section*{Acknowledgement}

This work is supported in part by the National Key R\&D Program of China under Grant No.
2023YFA1607104 and 2021YFC2203100, and the National Science Foundation of China under Grant No.~12475107.



\setcounter{equation}{0}
\counterwithout{equation}{section}
\setcounter{figure}{0}
\setcounter{table}{0}
\setcounter{section}{0}

\makeatletter
\renewcommand{\theequation}{A\arabic{equation}}
\renewcommand{\thefigure}{A\arabic{figure}}
\renewcommand{\thetable}{A\arabic{table}}

\appendix

\bibliographystyle{utphys}
\bibliography{refs}

\end{document}